\documentclass[11pt,a4paper]{article}

\usepackage{moreverb,url}

\usepackage[colorlinks,
            bookmarksopen,
            bookmarksnumbered,
            citecolor=red,
            urlcolor=red]{hyperref}

\newcommand\BibTeX{{\rmfamily B\kern-.05em \textsc{i\kern-.025em b}\kern-.08em
T\kern-.1667em\lower.7ex\hbox{E}\kern-.125emX}}

\usepackage[utf8]{inputenc}
\usepackage{bm}
\usepackage{amssymb,amsmath,amsthm}
\usepackage{mathtools}
\usepackage{graphicx}
\usepackage{hyperref}
\usepackage{booktabs}
\usepackage{multirow}
\usepackage{subfig}
\usepackage{authblk}

\usepackage{subfiles}

\DeclareMathOperator*{\argmin}{arg\,min}

\DeclarePairedDelimiter{\floor}{\lfloor}{\rfloor}

\begin{document}

\title{Robust Identification of Target Genes and Outliers in Triple-negative Breast Cancer Data}

\author[1]{Pieter Segaert}
\author[2]{Marta B. Lopes\footnote{The first two authors contributed equally to this work.}}
\author[3]{Sandra Casimiro}
\author[2,4]{Susana Vinga}
\author[1]{Peter J. Rousseeuw}

\affil[1]{Department of Mathematics, KU Leuven, Belgium}
\affil[2]{IDMEC, Instituto de Engenharia Mec\^anica, Instituto Superior T\'{e}cnico, Universidade de Lisboa}
\affil[3]{Lu\'is Costa Lab, Instituto de Medicina Molecular, Faculdade de Medicina da Universidade de Lisboa}
\affil[4]{INESC-ID, Instituto de Engenharia de Sistemas e Computadores - Investiga\c c\~ao e Desenvolvimento}

\date{}
\maketitle

\begin{abstract}
Correct classification of breast cancer sub-types is of high importance as it directly affects the therapeutic options. We focus on triple-negative breast cancer (TNBC) which has the worst prognosis among breast cancer types. Using cutting edge methods from the field of robust statistics, we analyze Breast Invasive Carcinoma (BRCA) transcriptomic data publicly available from The Cancer Genome Atlas (TCGA) data portal. Our analysis identifies statistical outliers that may correspond to misdiagnosed patients.  Furthermore, it is illustrated that classical statistical methods may fail in the presence of these outliers, prompting the need for robust statistics. Using robust sparse logistic regression we obtain 36 relevant genes, of which ca. 60\% have been previously reported as biologically relevant to TNBC, reinforcing the validity of the method. The remaining 14 genes identified are new potential biomarkers for TNBC. Out of these, \textit{JAM3}, \textit{SFT2D2} and \textit{PAPSS1} were previously associated to breast tumors or other types of cancer. The relevance of these genes is confirmed by the new DetectDeviatingCells (DDC) outlier detection technique. A comparison of gene networks on the selected genes showed significant differences between TNBC and non-TNBC data. The individual role of \textit{FOXA1} in TNBC and non-TNBC, and the strong \textit{FOXA1}-\textit{AGR2} connection in TNBC stand out. Not only will our results contribute to the breast cancer/TNBC understanding and ultimately its management, they also show that robust regression and outlier detection constitute key strategies to cope with high-dimensional clinical data such as omics data.
\end{abstract}

\section{Introduction}

Triple-negative breast cancer (TNBC) represents 10\% to 17\% of all diagnosed breast tumors \cite{Reis-Filho&Tutt_2008}, and has the worst prognosis amongst the different sub-types of breast cancer (BC) \cite{Dent_et_al_2007}. 
TNBC is characterized by the lack of expression of targetable genes like estrogen receptor (ER), progesterone receptor (PR) and human epidermal growth factor receptor 2 (HER2) \cite{Foulkes_et_al_2010}. Based on this, the current standard of care treatment protocols for TNBC are limited to surgery, radiotherapy and chemotherapy \cite{Wang_et_al_2017}. 
Since the BC sub-type directly influences the therapeutic options, there is a high demand for the development of methods that not only accurately classify BC patients into BC sub-types, but also identify relevant (target) genes that discriminate between TNBC patients and patients with other types of BC. The identification of genes that are either down- or up-regulated in TNBC is expected to play an important role in precision medicine, by providing a more in-depth knowledge on the cancer biology, but also yielding diagnostic, prognostic, and therapeutic markers that will ultimately improve patient outcomes \cite{Vucic_et_al_2012}.

The classification of BC into TNBC and non-TNBC is dependent on the presence of ER, PR and HER2 receptors, either `positive' or `negative',
based on the results obtained by immunohistochemical (IHC) and fluorescence in-situ
hybridization (FISH) testing technologies. However, preanalytic variables, thresholds for positivity, and interpretation criteria may generate inaccurate results. For example, it has been reported that up to 20\% of ER and PR test results are false negative or false positive \cite{Hammond_et_al_2010}. The clinical consequences are extremely important. A patient given a wrong BC subtype classification will undergo inappropriate cancer treatment, either hormonal based or not, with severe consequences for cancer progression and survival. False negatives for ER and PR will not benefit from endocrine therapy, and for false positives the hormonal therapy will fail. On the other hand, while a false positive HER2 assessment leads to the administration of potentially toxic, costly and ineffective HER2-targeted therapy, a false negative HER2 assessment results in denial of anti-HER2 targeted therapy for a patient who could benefit from it \cite{Wolff_et_al_2013}.

In this context, statistical analysis of gene expression data for known BC cases may provide valuable insights. However, real data often contain one or more observations deviating from the main pattern of the data \cite{Rousseeuw:RobReg,Maronna:RobStat}. For example, when considering gene expressions from TNBC data, inaccuracies may be due to variations in ER, PR and HER2 testing, as mentioned above. Wrong TNBC class labels may result from inconsistencies between the IHC and FISH testing technologies. 
Unfortunately, classical results are heavily influenced by these suspicious observations. The effect of outliers may be such that classical statistical techniques no longer detect them. This phenomenon is known as masking in statistics literature \cite{Rousseeuw:Unmasking}. Moreover, outlying observations may even influence classical statistics so much that regular observations are flagged as outliers, a phenomenon known as swamping (see e.g. \cite{serfling2014:maskingswamping}). In regression models, these observations may compromise the predictive performance. Due to the high dimensional nature of the data, typical regression techniques are no longer valid. For example, the classical least squares fit cannot be computed when there are fewer observations than variables. Therefore, one uses sparsity-inducing methods to select relevant subsets of the original variables. However, these sparse methods might also be impacted by outliers, leading to relevant variables being neglected and irrelevant variables being selected\cite{alfons2013}. Moreover, detecting interesting anomalous cases (e.g., a normal patient with deviating expressions of specific genes) may be of particular interest. 

The importance of detecting outlying patients is therefore twofold. Outliers corresponding to errors in the labeling must be detected and treated accordingly in order to achieve accurate model predictions. Correctly diagnosing patients is of utmost interest as wrongly diagnosed patients may receive ineffective, expensive and potentially harmful treatment. Secondly, outliers which are not errors reveal hidden information on the covariates that might play a role in the definition of new therapies based on target genes revealed by outlier analysis. 

The remainder of the paper is structured as follows. In the next section we dicuss TNBC data construction from RNA-Seq and clinical data from Breast Invasive Carcinoma (BRCA). We then discuss logistic regression as a tool to decide BC class membership. Due to the high dimensionality of the data and the concern for outliers, we then turn to robust sparse logistic regression which selects relevant variables and flags outlying cases. Also the \textit{DetectDeviatingCells} (DDC) method \cite{Rousseeuw:DDC} is applied, and its results are linked to those of the robust logistic regression which brings new insights. The paper concludes with a discussion of results and model diagnostics along with the biological interpretation of the selected gene set.

\section{Data description}
\label{sec:DataDescription}

The BRCA data set is publicly available from the Cancer Genome Atlas (TCGA) Data Portal \cite{TCGA} and contains genomic and clinical data from breast cancer (BC) patients. The RNA-Seq Fragments Per Kilo base per Million (FPKM) data was imported using the ‘brca.data’ R package \cite{BRCA_package}. The BRCA gene expression data is composed of 57251 variables for a total of 1222 samples from 1097 individuals. From those samples, 1102 correspond to primary solid tumor, 7 to metastases and 113 to normal breast tissue. Only samples from primary solid tumor were considered for analysis. 
	
The TNBC gene expression data set was built based on the BRCA RNA-Seq data set available from TCGA. A subset of 19688 variables, including the three TNBC-associated genes ER (ENSG00000091831), PR (ENSG00000082175) and HER2 (ENSG00000141736), was considered, corresponding to the protein coding genes reported from the Ensemble genome browser \cite{Ensembl} and the Consensus CDS \cite{CCDS} project. The response variable $Y$, corresponding to the clinical type, is a binary vector coded with `1' for TNBC individuals and `0' for non-TNBC. This vector was built based on the BRCA clinical data available from TCGA, regarding the individuals' label for ER, PR and HER2 (either  `positive',`negative' or `indeterminate'). When a `negative' label is recorded for all three genes the response is set to `1' (TNBC), whereas in all other cases the \textit{status} is `0' (non-TNBC). However, for assessing the HER2 label, three variables are available from the clinical data: two from the IHC analysis, the HER2 \textit{level} and \textit{status}, and another corresponding to the HER2 \textit{status} obtained by FISH. The IHC \textit{status} was considered, since it was measured for a larger number of individuals. Whenever both IHC \textit{status} and FISH \textit{status} were available for a given patient, the FISH \textit{status} was considered instead, as FISH is a more accurate test for classifying individuals into HER2 `positive’ or `negative’.

A total number of 28 individuals were marked as suspect when no concordance existed between the HER2 IHC \textit{level} and \textit{status}, and between the HER2 IHC \textit{status} and FISH \textit{status}. Special attention will be given to individuals for which non-concordance between lab testing exists and the choice of one or another determines the final label (TNBC vs. non-TNBC). These suspect individuals are potentially mislabeled, therefore potential outliers. We will verify whether they belong to the list of outlying individuals detected in this study. The variables age and race were also included as explanatory variables, since they were statistically significant in separate univariate logistic regressions to predict the individuals' \textit{status}.

\section{Methods}

Let $\bm{X} \in \mathbb{R}^{n \times p}$ denote the matrix of the predictors and $\bm{Y} \in \mathbb{R}^{n \times 1}$ the response vector. The logistic model may be formalized as \[\bm{Y} = \pi(\bm{X}) + \bm{\varepsilon} \] with \[\pi(\bm{X}) = \frac{\exp\left(\bm{X}\bm{\beta}\right)}{1 + \exp\left(\bm{X}\bm{\beta}\right)} \] and $\bm{\beta} = (\beta_1, \ldots, \beta_p)^t$ the vector of length $p$ containing the regression coefficients and $n$ the sample size. 

Typically, the regression coefficients $\bm{\beta}$ are estimated using the maximum likelihood estimator which minimizes the negative log-likelihood function: \[\hat{\bm{\beta}}_{ML} = \argmin_{\bm{\beta}} \sum_{i = 1}^{n}{d(\bm{x}_i, y_i; \bm{\beta})} \] where the {\it deviance} is $d(\bm{x}_i, y_i; \bm{\beta}) = \log\left(1 + \exp\left( \bm{x}_i^t\bm{\beta} \right)  \right) -y_i \bm{x}_i^t\bm{\beta} $.

However, when the number of variables $p$ is large, standard maximum likelihood estimators can be very difficult to interpret. Also the predictive power of a model may be impacted when including too many variables \cite{Breiman:SubsetSelection, Tibshirani_1996}.  Moreover, when $p > n$ the maximum likelihood estimator cannot even be computed. A possible solution to this problem is to consider so-called shrinkage estimators (for a review see e.g. Tibshirani, 2011\cite{Tibshirani:LASSORetro}). For these estimators, a penalty term on the regression coefficients is included in the objective function. In the next section we will discuss several shrinkage estimators. 

\subsection{Sparse logistic regression}

One of the first shrinkage estimators in the literature is ridge regression  \cite{Hoerl:RidgeBiased,Cessie:RidgeLogistic}.  The estimator for $\bm{\beta}$ then becomes the vector $\hat{\bm{\beta}}_{ridge}$ minimizing 
\[ \sum_{i = 1}^{n}{d(\bm{x}_i^t, y_i; \bm{\beta})} + \lambda ||\bm{p}^t \cdot \bm{\beta}||_2^2 = \sum_{i = 1}^{n}{d(\bm{x}_i^t, y_i; \bm{\beta})} + n\lambda \sum_{j = 1}^p \left( p_j \beta_j \right)^2. \] 
Here $\cdot$ stands for the elementwise product. The tuning parameter $\lambda > 0$ controls the severity of the penalty and thus the level of shrinkage. A higher value of $\lambda$ will lead to a higher importance of the penalty and thus a higher percentage of coefficients pulled towards zero. The vector $\bm{p}$ is a vector containing penalty factors that control how much of the penalty $\lambda$ affects each coefficient. If the $j$-th component of $\bm{p}$ is zero, the coefficient $\beta_j$ is not penalized. On the other hand if $p_j = 1$, $\beta_j$ the full penalty $\lambda$ is applied to $\beta_j$. 

A downside of ridge regression is that it cannot shrink coefficients exactly to zero. Ridge regression will keep all the variables in the model \cite{Breiman:SubsetSelection, Tibshirani_1996, Fan:VariableSelection}. The LASSO estimator \cite{Tibshirani_1996}, which employs an $l_1$ penalty instead of the $l_2$, may be used to solve this problem. It performs variable shrinkage and variable selection at the same time.  The LASSO-estimate of $\bm{\beta}$ is the vector $\hat{\bm{\beta}}_{LASSO}$ minimizing \[\sum_{i = 1}^{n}{d(\bm{x}_i^t, y_i; \bm{\beta})} + \lambda ||\bm{p} \cdot \bm{\beta}||_1 = \sum_{i = 1}^{n}{d(\bm{x}_i^t, y_i; \bm{\beta})} + \lambda \sum_{j = 1}^p |p_j \beta_j|. \] Again the tuning parameter $\lambda > 0$ controls the sparsity of the resulting coefficients. A downside of the LASSO estimator is that it tends to radomly select only one variable in a group of highly correlated variables while discarding the other variables \cite{Zou&Hastie_2005}.

The elastic net procedure proposed by Zou and Hastie (2005)\cite{Zou&Hastie_2005} tries to overcome the downsides of both ridge regression and the LASSO. It shrinks the variables and performs variable selection while being able to select groups of correlated variables. The sparse estimate for $\bm{\beta}$ then becomes the vector $\hat{\bm{\beta}}_{enet}$ minimizing
\[ \sum_{i = 1}^{n}{d(\bm{x}_i^t, y_i; \bm{\beta})} + n\lambda \left[(1-\alpha)\frac{||\bm{p} \cdot \bm{\beta}||_2^2}{2} + \alpha||\bm{p} \cdot \bm{\beta}||_1\right]. \] The  parameter $\alpha$ controls the mixing between the ridge and LASSO penalty and should be chosen in the interval $[0,1]$. Clearly, when $\alpha = 0$ the ridge estimator is obtained, whereas for $\alpha = 1$ the LASSO estimate is recovered. The optimal values for $\lambda$ and $\alpha$ may be obtained using $k$-fold cross-validation techniques. Software implementations of the elastic net method can be found in the free R software \cite{soft:R} package glmnet \cite{soft:glmnet}.

In the next subsection we discuss how the elastic net procedure may be modified to make it robust to outliers. We will first discuss a robustification of the non-sparse maximum likelihood technique before turning our attention a robust sparse procedure. 

\subsection{Robust sparse logistic regression}
The maximum likelihood estimator is highly susceptible to outliers, because both outliers in the predictor space and outliers in the response variable may have an unbounded effect on the log-likelihood. As a possible alternative, more robust procedures have been proposed. For an outlier-contaminated data set they provide a solution that is close to the one that would be obtained on an outlier-free data set using classical methods. One of the ways to achieve robustness is to use a trimmed log-likelihood function. For linear regression the resulting estimator is called the Least Trimmed Squares (LTS) estimator \cite{Rousseeuw:LTS, Rousseeuw:FASTLTS}.

For logistic regression the LTS estimator is defined by
  \[\hat{\bm{\beta}}_{LTS} = \argmin_{\bm{\beta}} 
  \sum_{i=1}^{h}{d(\bm{x}_i^t, y_i; \bm{\beta})_{i:n}} \]
where the subscript ${i:n}$ indicates the i-th smallest deviance, i.e. the $n$ deviances are first sorted from smallest to largest. The LTS thus minimizes the trimmed deviance, for a subset of $h$ data points out of the full sample of size $n$. The number $h$ is typically chosen between $\floor{(n+p+1)/2}$ and $n$. The choice of $h$ determines the robustness of the LTS estimator. In practice one frequently uses a conservative value of $h$ as an initial choice. To improve efficiency one may then increase $h$ to a higher value, while ensuring that $h$ stays below the number of non-outliers found in the data. 

The ideas of LTS regression were adapted for sparse robust logistic regression by Kurnaz and co-authors (2018) \cite{Kurnaz:RobustSparseLogistic} and were implemented in the R\cite{soft:R} software package enetLTS\cite{soft:enetLTS}. They defined the enet-LTS logistic estimator which combines the sparsity of the elastic net procedure with the robustness of LTS regression. In that sense their work is an extension of sparse LTS regression \cite{alfons2013} that combines the LTS estimator for linear regression with the LASSO penalty. 
The enet-LTS estimator is defined by
\[\hat{\bm{\beta}}_{\textit{enet-LTS}} = \argmin_{\bm{\beta}} 
\left( \sum_{i=1}^{h}{d(\bm{x}_i^t, y_i; \bm{\beta})_{i:n}} + 
h \lambda \left[(1-\alpha)\frac{||\bm{p} \cdot \bm{\beta}||_2^2}{2} +
\alpha||\bm{p} \cdot \bm{\beta}||_1\right] \right) \] 
where $\lambda \in [0, 1]$ as described for the glmnet penalty. 

To increase efficiency, LTS regression usually includes a reweighting step \cite{Rousseeuw:RobReg}. Generally speaking, the reweighting step identifies outliers according to the above fitted robust LTS model. These are then downweighted before fitting a classical model using these weights. Consider the Pearson residuals \[ r_i^s = \frac{y_i - \pi_i(\bm{x})}{\pi_i(\bm{x})(1-\pi_i(\bm{x}))}. \] Under the logistic model these are known to be approximately normally distributed. Each observation $i$ then receives a weight $w_i$ of 1 when $|r_i^s| < c$ and 0 otherwise. The cutoff value $c$ is determined as the $97.5$ quantile of a standard Gaussian distribution, so that $95\%$ of the distribution lies between $-1.96$ and $1.96\;$. The reweighted enet-LTS estimator  
is then defined as \[ \argmin_{\bm{\beta}} \left( \sum_{i = 1 }^{n}{w_i  d(\bm{x}_i^t, y_i; \bm{\beta})} + n_w \lambda \left[(1-\alpha)\frac{||\bm{p} \cdot \bm{\beta}||_2^2}{2} + \alpha||\bm{p} \cdot \bm{\beta}||_1\right] \right) \] where $n_w = \sum{w_i}$ is the number of observations receiving weight one.

\subsection{Detecting Deviating Data Cells}
Let $\bm{X} \in \mathbb{R}^{n \times p}$ now denote a data matrix of sample size $n$ and dimension $p$. In statistics, an outlier typically refers to a row (case) that deviates from the bulk of the data. However, it frequently occurs that such a row is only suspicious for $q$ out of the $p$ variables, with $q \ll p\;$
\cite{alqallaf2009propagation, van2012stahel, agostinelli2015robust, ollerer2016shooting, leung2016robust}. Flagging the entire row as an outlier would thus be too conservative as the remaining $p - q$ measurements of that row still contain valuable information. To work in this paradigm, we no longer see the data as $n$ rows of $p$ variables, but rather as a data matrix of $n \times p$ cells. Cells with possibly deviating behavior are then referred to as cellwise outliers. 

\begin{figure}[h!]
    \centering
    \includegraphics[width = 0.5\textwidth]{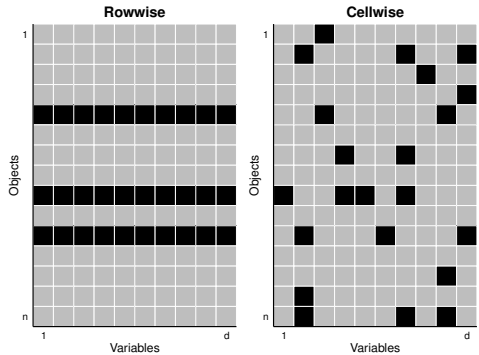}
    \caption{Illustration of the cellwise outlier paradigm versus the typical outlier paradigm.}
    \label{fig:cellWise_Illustration}
\end{figure}

Figure~\ref{fig:cellWise_Illustration} illustrates these two different paradigms. The left panel indicates three rowwise outliers in the data. The right panel identifies several contaminated cells in the data matrix. Even though many rows have one or more outlying cells, they still contain valuable information in their remaining cells. 

In the context of the TNBC data, one may expect that TNBC patients would only have deviating observations for a small subset of the full gene expression data. We will therefore analyze the genes selected by the robust sparse logistic regression using the \textit{DetectDeviatingCells} (DDC) method recently proposed by Rousseeuw and Van den Bossche (2018)\cite{Rousseeuw:DDC}. This will provide us with additional insight in the nature of flagged outliers and the role of the selected genes. 

The \textit{DetectDeviatingCells} procedure uses bivariate correlations between the different variables. It then computes a predicted value for each cell, based on the values of other cells in the same row. Next, it compares the predicted and observed value of each cell. When this robustly standardized difference exceeds a certain cutoff, the cell is flagged. Cells for which the observed value is much lower then the imputed value are colored blue. When the observed value is higher than the imputed value, the cell is colored red.

\section{Results and discussion}

We analyzed the TNBC data set using both the sparse logistic and the robust sparse logistic procedures discussed above. In both instances, the parameters $\alpha$ and $\lambda$ were selected using $5$-fold cross-validation evaluating the mean of the deviances in the fold. To eliminate randomness in the selection of the folds, the cross-validation was averaged over 10 runs. For the robust sparse logistic regression method, the parameter $h$ was selected as $0.85 n$. This parameter was found to be a safe level guarding against outliers after an initial run with $h = 0.5 n$. The penalty factor $\bm{p}$ was chosen to be a unity vector penalizing all coefficients equally, except for the coefficients of ER, PR and HER2 for which $p_j = 0.5$ as these are of special interest in the TNBC context.

Table \ref{tab:basicResults} summarizes the sparse fits. For both procedures, we list the selected $\alpha$ and $\lambda$ parameters in the glmnet procedure and the resulting number of non-zero coefficients. We also provide the number of observations that are flagged as outliers with respect to the fitted model. The criterion used to flag outliers in the classical sparse logistic model corresponds to the procedure described in the reweighting step of the robust sparse logistic regression method.

\begin{table}[htb]
\centering
\caption{Summary of the fitted models for the robust and non-robust sparse logistic regression methods}
\label{tab:basicResults}
\begin{tabular}{l|ll}
\toprule
                      & sparse & robust sparse\\ 
                      & logistic regression & logistic regression \\ 
\toprule
$\alpha$              & 1.00                        & 0.81                               \\
$\lambda$             & 0.005                      & 0.057                             \\
$\#$ of non-zero coefficients & 136                         & 36                               \\ 
\midrule
Potential outliers    & 0                         & 43                                \\ 
\bottomrule
\end{tabular}
\end{table}

The results in Table \ref{tab:basicResults} show that the cross-validation leads to different parameter choices for $\alpha$ and $\lambda$ between the robust and non-robust sparse logistic regression method. The classical method selects roughly four times as many coefficients (genes) as the robust method. Moreover, only three of the 136 genes selected by the classical method are also selected by the robust method, namely \textit{ER}, \textit{HER2} and \textit{PPP1R14C}. 
While the nonrobust method fails to identify outlying observations, due to the masking effect described in the introduction, the robust procedure indicates 43 observations as potential outliers. It is important to note that all 43 outliers flagged by the robust method are marked as non-TNBC patients in the data, but are predicted to be TNBC according to the fitted model. If indeed true, these patients would receive potentially toxic, costly and ineffective therapies. It is therefore important to detect all such cases. 

As neither model selects the variables age and race, the 95 observations that were initially left out of the analysis due to missing values for either of these variables may be used for out-of-sample testing. The prediction of TNBC occurrence does not match with the data in 4 out of 95 cases for the classical method and in 6 of the 95 cases for the robust method. Both methods find two additional outliers in the test set corresponding to patients who were attributed a non-TNBC status in the data. From the 6 cases who did not match according to the robust method, two were borderline cases with a predicted TNBC likelihood of 0.53 and 0.57 respectively. Even though the robust model selects only a handful of genes compared to the classical method, its out of sample performance is comparable to traditional methods.

\subsection{Detailed discussion of identified outliers}

\begin{table}[htbp]
\centering
\scriptsize
\caption{Summary of the 43 individuals identified as outliers by robust sparse logistic regression regarding ER, PR and HER2 gene expression and corresponding clinical label (between brackets). Individuals highlighted in bold correspond to individuals previously identified as suspicious as described in the section "Data description". (We abbreviate fragments per kilobase million by FPKM, indeterminate by Ind and equivocal by Equiv).
}
\label{tab:gene_expression_outliers}
\begin{tabular}{rcccccccc}
\hline
& & \multicolumn{6}{c}{Genes} & \\
\cline{3-8}\\
& Individual & ER & PR & \multicolumn{4}{c}{HER2} & Clinical type \\
\cline{5-8}\\
&  & FPKM & FPKM & FPKM & \multicolumn{3}{c}{(clinical)} & \\
\cline{6-8}
& & (clinical) & (clinical) & & \textit{level} & \textit{status} & \textit{status} &  \\
& & & & & IHC & IHC & FISH & \\
\hline
1 & TCGA-AR-A1AO & 1.47(+) & 1.13(-) & 14.89 & (-) & (-) &  & non-TNBC \\
2 & TCGA-BH-A6R9 & 0.59(-)      & 0.25(+)      & 8.18   &              & (-)      &   & non-TNBC \\
3 & TCGA-AC-A62X & 0.19(+)      & 0.02(-)      & 28.53  &              &        &   & non-TNBC \\
4 & TCGA-A2-A0YJ & 0.09(+)      & 0.03(-)      & 240.24 & (-)            & (-)      &   & non-TNBC \\
5 & \textbf{TCGA-LL-A5YP} & \textbf{0.16(+)} & \textbf{0.05(-)} & \textbf{15.10}  & (\textbf{-}) & (\textbf{-}) & (\textbf{+}) & \textbf{non-TNBC} \\
6 & TCGA-A7-A13D & 0.52(-)      & 0.81(+)      & 42.28  & (Ind)          & (Equiv)  & (-) & non-TNBC \\
7 & TCGA-E2-A1II & 0.14(-)      & 0.19(+)      & 10.73  & (-)            & (-)      &   & non-TNBC \\
8 & TCGA-AR-A1AH & 0.03(+)      & 0.03(-)      & 34.12  &              & (-)      &   & non-TNBC \\
9 & TCGA-BH-A0DL & 6.99(+)      & 0.04(-)      & 9.92   &              & (-)      &   & non-TNBC \\
10 & TCGA-E2-A14Y & 0.67(+)      & 0.03(+)      & 487.90 & (Ind)          & (Equiv)  & (+) & non-TNBC \\
11 & \textbf{TCGA-AO-A0JL} & \textbf{0.63(-)} & \textbf{0.08(-)} & \textbf{63.60} & (\textbf{-}) & (\textbf{-}) & (\textbf{+}) & \textbf{non-TNBC} \\
12 & \textbf{TCGA-AN-A0FL} & \textbf{0.09(-)} & \textbf{1.07(-)} & \textbf{15.07} & (\textbf{-}) & (\textbf{+}) &  & \textbf{non-TNBC} \\
13 & TCGA-AO-A1KO & 10.78(+)     & 9.12(+)      & 14.91  & (-)            & (-)      &   & non-TNBC \\
14 & \textbf{TCGA-AN-A0FX} & \textbf{1.13(-)} & \textbf{0.64(-)} & \textbf{24.02} & (\textbf{-}) & (\textbf{+}) &  & \textbf{non-TNBC} \\
15 & TCGA-A1-A0SB & 3.16(+)      & 0.03(-)      & 32.35  &              & (-)      &   & non-TNBC \\
16 & TCGA-D8-A1JM & 5.01(+)      & 0.01(-)      & 21.85  & (-)            & (-)      &   & non-TNBC \\
17 & TCGA-E9-A1NC & 0.11(-)      & 0.08(+)      & 15.91  &              & (+)      &   & non-TNBC \\
18 & TCGA-A2-A25F & 0.62(-)      & 0.23(+)      & 5.19   &              & (-)      &   & non-TNBC \\
19 & TCGA-A2-A1G1 & 0.53(-)      & 0.17(-)      & 819.76 & (Ind)          & (Equiv)  & (+) & non-TNBC \\
20 & TCGA-LL-A6FR & 0.33(-)      & 0.04(+)      & 32.13  & (Ind)          & (Equiv)  & (+) & non-TNBC \\
21 & TCGA-A2-A3Y0 & 2.18(+)      & 0.03(-)      & 11.34  & (-)            & (-)      &   & non-TNBC \\
22 & TCGA-B6-A0IJ & 1.18(+)      & 0.46(+)      & 11.12  &              &        &   & non-TNBC \\
23 & TCGA-AR-A0TP & 0.04(+)      & 0.03(-)      & 13.39  &              & (-)      &   & non-TNBC \\
24 & TCGA-S3-AA0Z & 16.67(+)     & 0.07(+)      & 33.07  & (-)            & (Equiv)  & (-) & non-TNBC \\
25 & TCGA-A2-A4S1 & 0.29(+)      & 0.01(-)      & 0.61   &              & (-)      &   & non-TNBC \\
26 & TCGA-A7-A13E & 0.82(+)      & 0.06(-)      & 46.08  & (Ind)          & (Equiv)  & (-) & non-TNBC \\
27 & TCGA-D8-A1JK & 0.40(-)      & 0.72(+)      & 22.19  & (-)            & (-)      &   & non-TNBC \\
28 & TCGA-E9-A1ND & 1.44(-)      & 0.05(-)      & 13.05  &              & (+)      &   & non-TNBC \\
29 & \textbf{TCGA-JL-A3YW} & \textbf{0.35(+)} & \textbf{0.09(+)} & \textbf{31.47} & \textbf{(-)} & \textbf{(+)} &  & \textbf{non-TNBC} \\
30 & \textbf{TCGA-AN-A0FJ} & \textbf{0.08(+)} & \textbf{0.04(-)} & \textbf{14.28} & (\textbf{-}) & (\textbf{+}) &  & \textbf{non-TNBC} \\
31 & TCGA-D8-A1XW & 0.32(-)      & 0.11(+)      & 21.03  & (-)            & (-)      &   & non-TNBC \\
32 & TCGA-UU-A93S & 0.30(-) & 0.12(-) & 1668.35 & (+) & (+) &  & non-TNBC \\
33 & TCGA-OL-A5S0 & 0.09(+)      & 0.06(-)      & 31.92  &              &        & (+) & non-TNBC \\
34 & TCGA-E9-A22G & 0.44(-)      & 0.02(-)      & 15.32  &              & (+)      &   & non-TNBC \\
35 & TCGA-AR-A24Q & 1.00(+)      & 0.36(-)      & 20.67  &              & (-)      &   & non-TNBC \\
36 & TCGA-E2-A1B0 & 0.14(-)      & 0.26(-)      & 563.81 & (+)            & (+)      &   & non-TNBC \\
37 & TCGA-AR-A251 & 1.57(+)      & 0.10(-)      & 14.02  & (Ind)          & (Equiv)  & (-) & non-TNBC \\
38 & TCGA-A2-A4RX & 0.68(+)      & 0.93(+)      & 26.64  & (-)            & (-)      &   & non-TNBC \\
39 & TCGA-AR-A1AJ & 1.47(+)      & 0.07(-)      & 9.74   &              & (-)      &   & non-TNBC \\
40 & \textbf{TCGA-A2-A04U} & \textbf{0.02(-)} & \textbf{0.02(-)} & \textbf{9.64} & (\textbf{-}) & (\textbf{-}) & (\textbf{+}) & \textbf{non-TNBC} \\
41 & TCGA-BH-A5IZ & 5.12(+)      & 0.03(-)      & 28.08  &              & (-)      & (-) & non-TNBC \\
42 & TCGA-D8-A13Y & 15.48(+)     & 4.17(+)      & 4.83   & (-)            & (-)      &   & non-TNBC \\
43 & TCGA-LL-A8F5 & 1.08(+)      & 0.04(-)      & 11.86  & (-)            & (-)      &   & non-TNBC \\
\hline
\end{tabular}
\end{table}

We now turn our attention to the observations flagged as potential outliers by the robust method and and investigate how these observations differ from the others. The expression of ER, PR and HER2 for the flagged outliers, along with their label (TNBC vs. non-TNBC), can be found in Table \ref{tab:gene_expression_outliers}. 

All observations flagged as outliers are originally labeled non-TNBC individuals.
From the list of 43 outliers detected, 7 were previously identified as suspect, i.e., individuals for which there is no concordance between lab methods for the HER2 determination (see Section Info on the Data). This is particularly critical for individuals ER and PR `negative' and for which the HER2 label determines the final TNBC vs. non-TNBC label. For instance, if the other HER2 (correct) testing result was chosen in individuals 12, 34 and 40, the final label would have been TNBC.

Several inconsistencies in labeling can be also observed for ER and PR testing. An ER expression value of 0.03, for example, has corresponding positive label for individual 8, while an ER value of 0.63 is translated into a negative label for individual 11. Regarding PR labeling, the same PR expression value of 0.03 corresponds to a negative label for individual 8 and to a positive label for individual 10. Similarly, individual 2 with a PR expression value of 0.25 has a positive PR label, while individual 36 with a PR expression value of 0.26 has a negative PR label. Besides uncertainty in HER2 labeling with respect to the different testing, inconsistencies can be also observed within a given test, e.g., the \textit{status} by IHC. Individual 5 was labeled negative with a HER2 expression value of 15.1, and for the same expression value individual 12 was labeled as HER2 positive. Wrong labeling in one or more variables clearly impacts final labeling of individuals into TNBC and non-TNBC, with serious consequences in clinical decision and prognosis.

For other individuals, however, the outlyingness cannot be explained by mislabeling of the three TNBC-associated gene expressions, as they seem to have concordant gene expression and label (see e.g. individuals 9, 37 and 42). This suggests that other genes than ER, PR and HER2 might contribute to the discrimination between TNBC and non-TNBC individuals.

\begin{figure}[htbb]
    \centering
    \includegraphics[width = 0.9\textwidth, angle = -90]{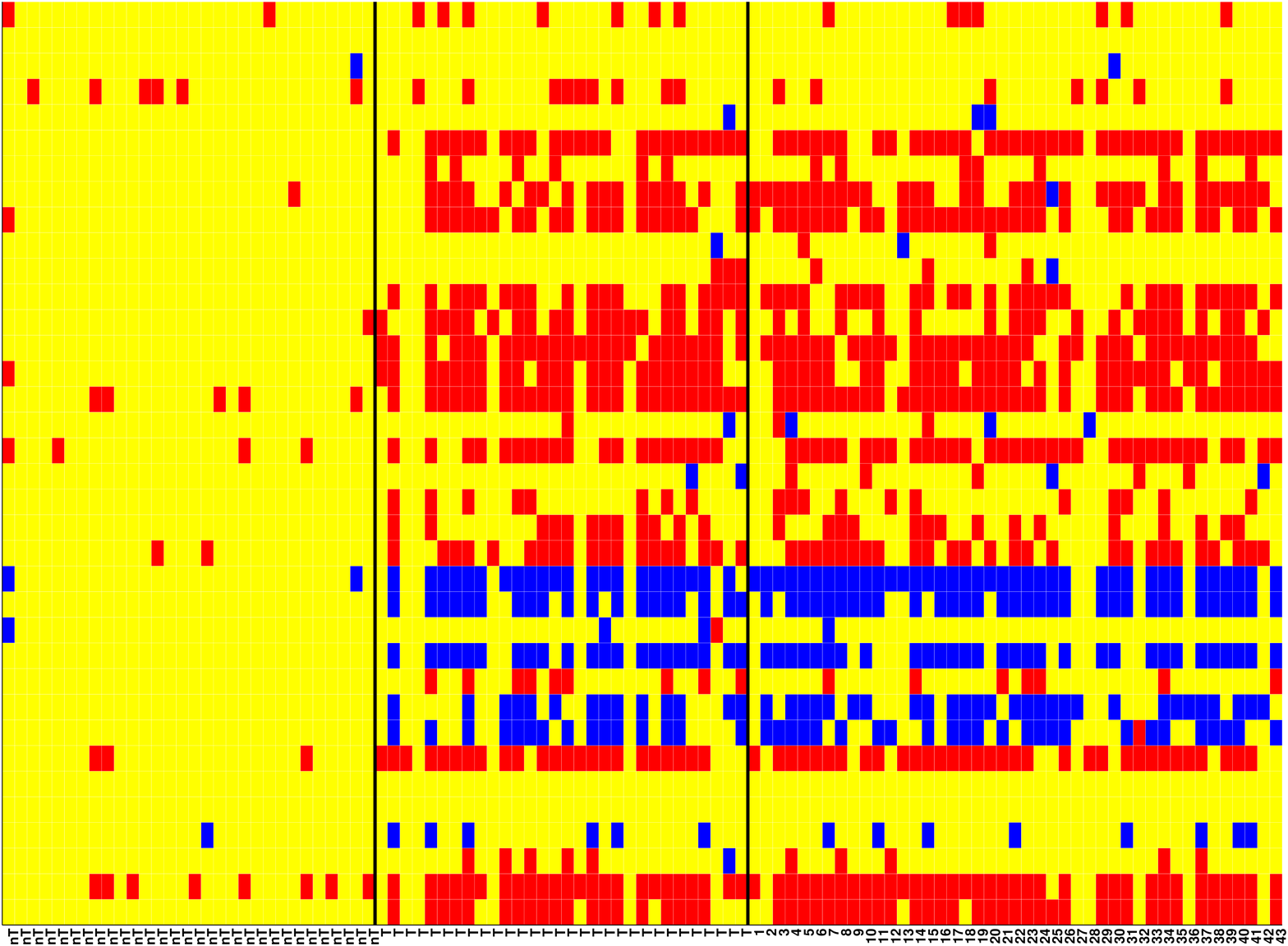}
    \caption{Cellwise outlier map. The columns correspond to the genes selected by the robust sparse logistic model. The rows correspond to 30 non-TNBC patients (label nT), 30 TNBC patients (label T) and the 43 outliers found by the robust fit.}
    \label{fig:cellMap}
\end{figure}

\subsection{Discussion of selected genes with relation to outlier identification}

To gain more insight we ran the DetectDeviatingCells (DDC) algorithm on the 36 selected genes only, without telling DDC anything about the clinical response variable or which rows were flagged as outliers. The result is a cell map with over 1000 rows, which is hard to visualize. Therefore Figure \ref{fig:cellMap} instead shows (from top to bottom) the first 30 non-TNBC patients, 30 TNBC patients, and the 43 outliers found by the robust logistic fit. The indices of the outliers correspond to the row numbers in Table \ref{tab:gene_expression_outliers}. A blue cell in Figure \ref{fig:cellMap} indicates an unexpectedly low gene expression value whereas a red cell indicates an unexpectedly high value, relative to the other cells in its row and using the correlations between the columns. We see that the overall pattern detected by the DDC algorithm for the patients flagged as potential outliers (all originally labeled as non-TNBC) matches the pattern observed for the TNBC patients. This is a very strong indication that indeed these patients have an erroneous label in the data.

\begin{figure}[tb]
  \centering
\includegraphics[width = 0.85\textwidth]{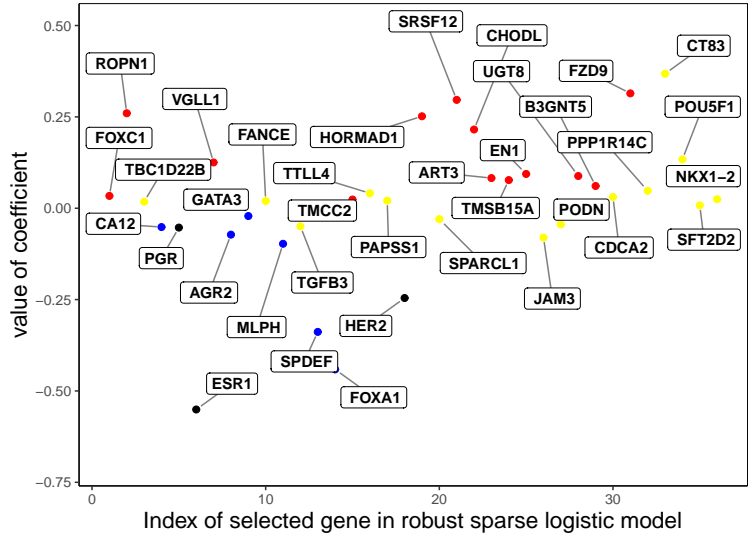}
  \caption{Interpretation of genes selected in the robust sparse logistic model. The color coding corresponds to the color determined by the DDC map.}
    \label{fig:CoefGen}
 \end{figure}

Genes for which most of the cells of the TNBC patients are colored are of particular interest. We may verify their predictive power for the classification of TNBC patients by the size of the coefficients in the robust sparse logistic model. Figure \ref{fig:CoefGen} depicts the coefficient in the robust sparse logistic model for each of the genes, using the same color coding as in the DDC map. The coefficients of the ER, PR and HER2 gene receptors have been colored black. We indeed see that the genes standing out in the DDC map are mostly those genes with higher coefficients, in absolute value, in the model. The red-colored genes turn out to get positive coefficients, the blue genes get negative coefficients, and the yellow ones get coefficients closer to zero. The selected genes may thus be of particular biological and medical interest for the understanding and diagnosis of TNBC.

Table \ref{tab:selectedGenesSorted} lists the genes selected by the robust sparse logistic method. It also includes the corresponding coefficients estimated by the robust sparse logistic method, rounded to 3 digits. The color coding as determined by the DDC map is also noted and corresponds to the color coding in Figure \ref{fig:CoefGen}.

 \begin{table}[tb]
 \caption{Genes selected by the robust sparse logistic method, corresponding coefficients (rounded to 3 digits) and their color coding. The genes are sorted by their coefficient.}
 \label{tab:selectedGenesSorted}
 \centering
 \tiny

 \begingroup
 \renewcommand{\arraystretch}{0.95}

 \begin{tabular}{rlll|rlll}
   \hline
  & gene & coef & color & & gene & coef & color \\ 
   \hline
   0 &  Intercept & 0.225 & None &19&  \textit{TMCC2} & 0.024 & red \\ 
   1 &  \textit{CT83} & 0.368 & yellow &20&  \textit{PAPSS1} & 0.021 & yellow \\ 
   2 &  \textit{FZD9} & 0.314 & red &21&  \textit{FANCE} & 0.020 & yellow \\ 
   3 &  \textit{SRSF12} & 0.297 & red &22&  \textit{TBC1D22B} & 0.018 & yellow \\ 
   4 &  \textit{ROPN1} & 0.260 & red &23&  \textit{SFT2D2} & 0.008 & yellow \\ 
   5 &  \textit{HORMAD1} & 0.252 & red &24&  \textit{GATA3} & -0.021 & blue \\ 
   6 &  \textit{CHODL} & 0.215 & red &25&  \textit{SPARCL1} & -0.030 & yellow \\ 
   7 &  \textit{POU5F1} & 0.134 & yellow &26& \textit{PODN} & -0.044 & yellow \\ 
   8 &  \textit{VGLL1} & 0.125 & red &27& \textit{TGFB3} & -0.050 & yellow \\ 
   9 &  \textit{EN1} & 0.094 & red &28&  \textit{CA12} & -0.051 & blue \\ 
   10 &  \textit{UGT8} & 0.088 & red &29& \textit{PGR} & -0.053 & yellow \\ 
   11 &  \textit{ART3} & 0.083 & red &30&  \textit{AGR2} & -0.072 & blue \\ 
   12 &  \textit{TMSB15A} & 0.077 & red &31&  \textit{JAM3} & -0.080 & yellow \\ 
   13 & \textit{B3GNT5} & 0.061 & red &32&  \textit{MLPH} & -0.097 & blue \\ 
   14 &  \textit{PPP1R14C} & 0.048 & yellow &33& \textit{HER2} & -0.246 & yellow \\ 
   15 &  \textit{TTLL4} & 0.041 & yellow &34&  \textit{SPDEF} & -0.338 & blue \\ 
   16 &  \textit{FOXC1} & 0.034 & red &35&  \textit{FOXA1} & -0.441 & blue \\ 
   17 &  \textit{CDCA2} & 0.031 & yellow &36& \textit{ESR1} & -0.551 & yellow \\ 
   18 &  \textit{NKX1-2} & 0.025 & yellow &  &  &  \\ 
   \hline
 \end{tabular}
 \endgroup
 \end{table}

\subsection{Biological interpretation of selected genes and correlation structures}

Among the 36 genes listed in Figure \ref{fig:CoefGen}, 13 (36.1\%) were down-regulated in TNBC, and 23 (63.9\%) were up-regulated. The majority of genes found to be down-regulated in TNBC (11/13) were previously reported to be down-regulated in this particular sub-type of BC, or overexpressed in ER+/HER2+ breast tumors. These include \textit{ESR1}, \textit{PGR} and \textit{HER2}, but also \textit{AGR2} \cite{Tian_et_al_2017}, \textit{CA12} \cite{Christgen_et_al_2013}, \textit{FOXA1} \cite{Guiu_et_al_2015}, \textit{GATA3}\cite{Krings_et_al_2014,Cimino-Mathews_et_al_2013}, \textit{MLPH} \cite{Thakkar_et_al_2015,Thakkar_et_al_2010}, \textit{SPDEF} \cite{Turcotte_et_al_2007}, \textit{SPARCL1} \cite{Cao_et_al_2013}, and \textit{TGFB3}\cite{Chen_et_al_2015}. Also 11 of the 23 genes up-regulated in TNBC were previously described as such, namely \textit{ART3} \cite{Tan_et_al_2016}, \textit{B3GNT5} \cite{Potapenko_et_al_2015}, \textit{EN1} \cite{Beltran_et_al_2014, Kim_et_al_2018}, \textit{FOXC1} \cite{Jensen_et_al_2015}, \textit{FZD9} \cite{Conway_et_al_2014}, \textit{HORMAD1} \cite{Watkins_et_al_2015}, \textit{POU5F1} \cite{Liu_et_al_2011}, \textit{ROPN1} \cite{Ivanov_et_al_2013}, \textit{TMSB15A} \cite{Darb-Esfahani_et_al_2012}, \textit{UGT8} \cite{Dziegiel_et_al_2010}, and \textit{VGLL1} \cite{Castilla_et_al_2014}). 

Our analysis has led to the identification of 14 genes that were not previously reported as specifically involved in TNBC or (breast) cancer overall, therefore contributing to the search for new interest biomarkers to further validate and functionally study. These include \textit{JAM3} and \textit{PODN}, down-regulated in TNBC; and \textit{SFT2D2}, \textit{CDCA2}, \textit{CHODL}, \textit{CT83}, \textit{FANCE}, \textit{NKX1-2}, \textit{PPP1R14C}, \textit{SRSF12}, \textit{TBC1D22B}, \textit{TMCC2}, \textit{TTLL4}, and \textit{PAPSS1}, up-regulated in TNBC. \textit{JAM3} has been previously identified as a biomarker for cervical cancer \cite{Eijsink_et_al_2012}; and was found to be up-regulated and associated with higher cancer risk in the offspring from mice with high fat diet intake during pregnancy \cite{Nguyen_et_al_2017}. Amongst the up-regulated genes, \textit{SFT2D2} was previously described as down-regulated in a bone (specific) metastasis-related gene set \cite{Savci-Heijink_et_al_2016}. \textit{PAPSS1}, involved in estrogen metabolism, was not directly implicated in TNBC before but found to be overexpressed in breast tumor tissues in comparison to adjacent normal tissue, independently of ER status \cite{Xu_et_al_2012}.

\begin{figure}[tbp]
  \centering
  \caption{Representation of the correlation between the genes selected by the robust sparse logistic model. The color coding corresponds to the color determined by the DDC map.}
  \label{fig:CorrGen}

  \subfloat[Correlations for non-TNBC patients.]{\label{fig:CorrGen_nonTNBC}\includegraphics[height = 0.5\textwidth, angle = -90]{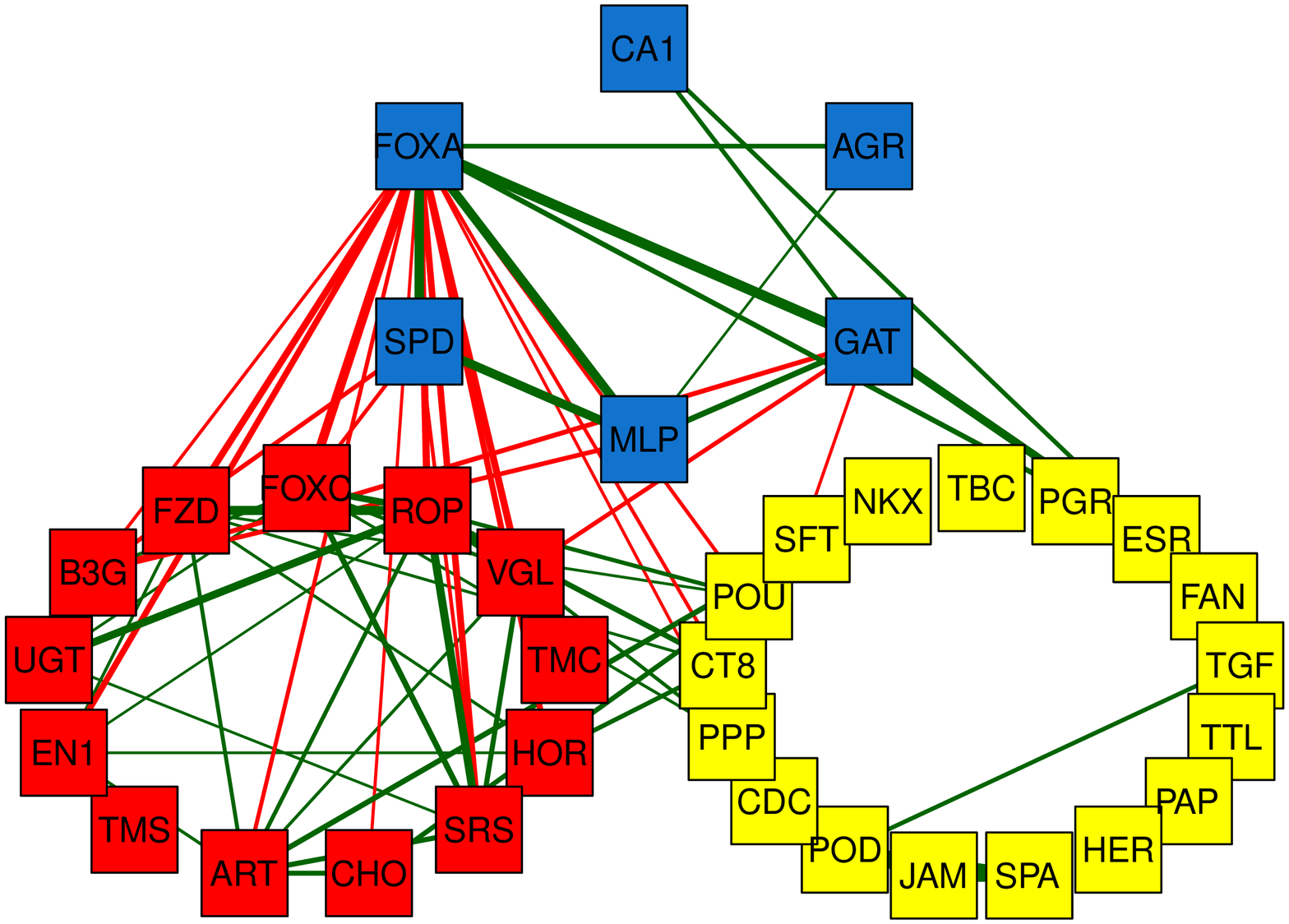}}
  \subfloat[Correlations for TNBC patients.]{\label{fig:CorrGen_TNBC}\includegraphics[height = 0.5\textwidth, angle = -90]{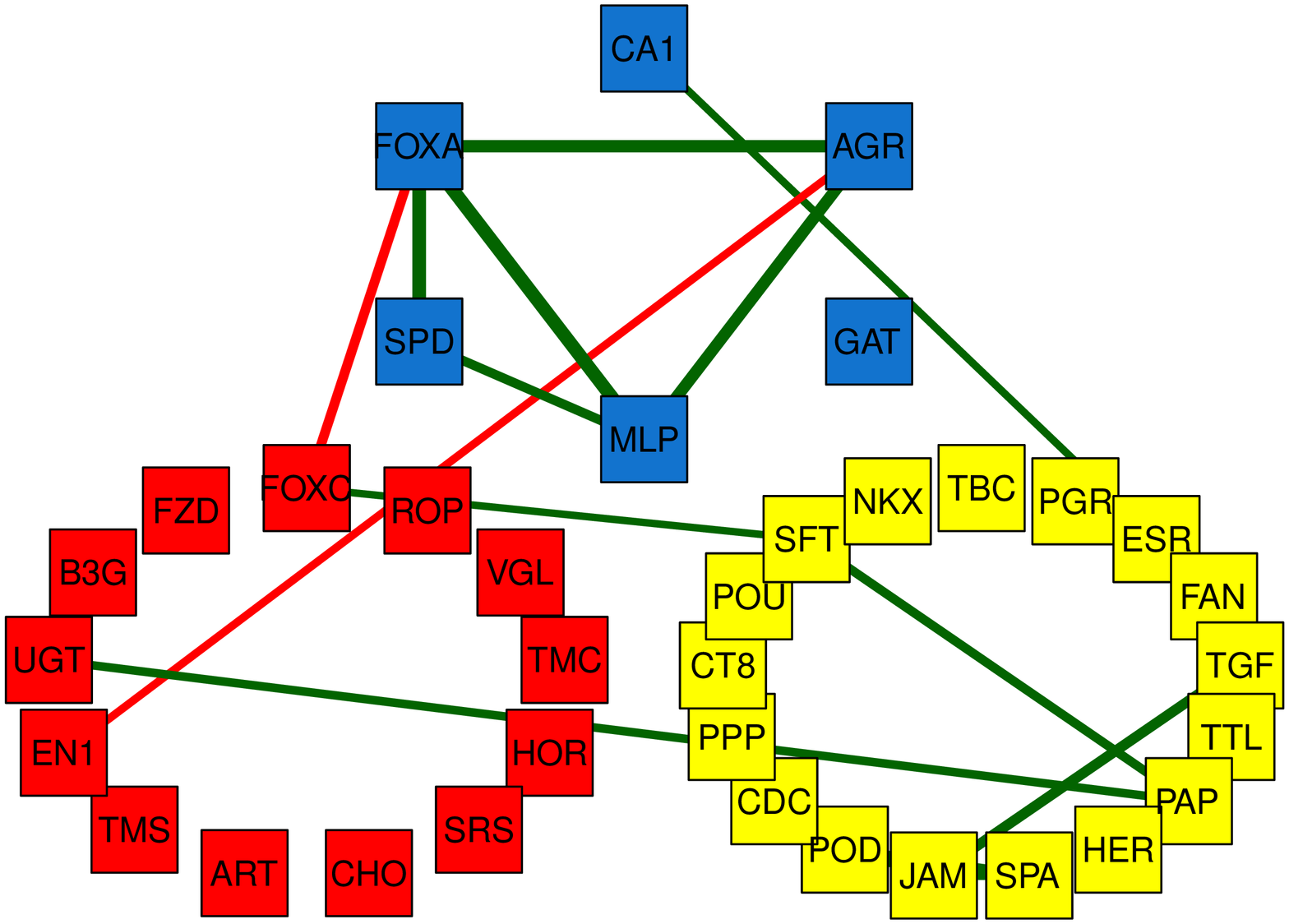}}
\end{figure}

Finally, we consider a graphical representation of the correlation-based network structure between the selected genes depicted in Figure \ref{fig:CorrGen}. For clarity, only correlations above 0.6 in absolute value are shown. The thickness of the connecting lines represents the strength of the correlation, whereas green represents a positive correlation and red signals a negative correlation. Figure \ref{fig:CorrGen_nonTNBC} shows the correlation plot for the non-TNBC individuals whereas Figure \ref{fig:CorrGen_TNBC} is the plot for TNBC individuals. The patterns in the left and right panels are strikingly different.

\vspace{2mm}

The proto-oncogene \textit{AGR2} (anterior gradient homology 2) is among the down-regulated genes in TNBC, and a strong correlation between \textit{AGR2} and \textit{FOXA1} was found in TNBC. Moreover, \textit{AGR2} was correlated with other genes, suggesting a particular relevance for this gene. \textit{AGR2} is a known biomarker of poor prognosis in ER+ BC \cite{Dong_et_al_2009}. Accordingly, different studies have reported that expression of the proto-oncogene \textit{AGR2} is induced by estrogen and tamoxifen in BC cells \cite{Hrstka_et_al_2010}; and that \textit{AGR2} is required for the growth and migration of ER+ cells. The transcription factor \textit{FOXA1} is implicated in the regulation of many ER$\alpha$-target genes, including \textit{AGR2}. This justifies the multiple correlations we found between \textit{FOXA1} and other genes in non-TNBC. 
However, in tamoxifen-resistant cells, the expression of \textit{AGR2} occurs in a constitutive manner, requiring \textit{FOXA1}, but loses its dependence on ER, suggesting a mechanism where changes in \textit{FOXA1} activity obviate the need for ER in the regulation of this gene \cite{Wright_et_al_2014}. It is hypothesized that \textit{AGR2} may be involved in the folding of the extracellular domains of proteins that influence cell growth and survival, and that \textit{AGR2} may represent an important biomarker and therapeutic target in BC \cite{Salmans_et_al_2013}. It thus appears that the \textit{FOXA1}-\textit{AGR2} link in TNBC may be of particular relevance and deserves further study.

The fact that the biological role in TNBC of approximately 60\% of the selected genes has been previously reported strengthens our analysis and fosters investigation on the potential role of the remaining selected genes in BC and in particular TNBC.

\subsection{Conclusion}

This work shows that robust sparse logistic regression can be a powerful tool in precision medicine. It enables accurate prediction of the BC subtype, irrespective of the possible presence of outliers situated in either the clinical label or in the gene expression data. In contrast, classical sparse logistic regression methods are severely affected by the outliers in the data. At the same time, robust methodology allows to inspect the detected outliers which may lead to the correct status of the patient and the prescription of the appropriate treatment. Due to the sparse nature of this robust regression, genes included in the model may be highly relevant to the understanding of TNBC.  While 60\% of the selected genes were previously reported to be related to TNBC or BC, the remaining identified genes deserve further attention as potential biomarkers for the disease. Among the selected genes, biologically relevant gene networks could be identified both for TNBC and non-TNBC patient data, particularly the strong \textit{FOXA1}-\textit{AGR2} link in TNBC. These results are intended to contribute to BC/TNBC understanding, the definition of new therapies and ultimately more effective TNBC management.

\subsection{Acknowledgments and Funding}
The authors thank Andr\'e Ver\'issimo and Eunice Carrasquinha for insightful discussions during problem identification and data processing.\\ 

This work was supported by the International Funds KU Leuven [grant C16/15/068]; the European Union Horizon 2020 research and innovation program [grant No. 633974 (SOUND project)], and the Portuguese Foundation for Science \& Technology (FCT), through projects UID/EMS/50022/2013 (IDMEC, LAETA), UID/CEC/50021/2013 (INESC-ID), PTDC/EMS-SIS/0642/2014 and IF/00653/2012. The computational resources and services used in this work were provided by the VSC (Flemish Supercomputer Center), funded by the Flemish Research Foundation (FWO) and the Flemish Government, department EWI.


\bibliographystyle{SageV}
\bibliography{refs}

\end{document}